\documentclass[12pt]{article} 
\usepackage[margin=1in]{geometry}
\usepackage{appendix}
\usepackage[english]{babel} 
\usepackage{amssymb}
\usepackage{float}
\usepackage{amsmath}
\usepackage{amsfonts}
\usepackage{txfonts}
\usepackage{mathdots}
\usepackage[classicReIm]{kpfonts}
\usepackage[final]{graphicx}
\graphicspath{ {./} }
\usepackage{url}
\begin{document}
\thispagestyle{empty}
\begin{titlepage}
\begin{center}
    \vspace*{0.5in}
    {\Large\bf Frequency Estimation of Gravitational Waves from a Binary Black Hole Merger} \\
    \vspace{0.6in}
    {\em Report submitted in fulfillment of the Internship}\\
    \vspace{0.2in}
    % {\Large Doctor Of Philosophy}\\
    \vspace{0.1in}
    {\em in}\\
    \vspace{0.1in}
    {\Large Electrical Engineering Department}\\
    \vspace{0.5in}
    {\em by}\\
    \vspace{0.5in}
    {\Large \bf VARUN NAGESH JOLLY BEHERA}\\
    \vspace{0.5in}
    {\em under the guidance of}\\
    \vspace{0.2in}
    {\bf Prof. Aurobinda Routray\\
    \bf Mr. Anik Kumar Samanta}\\
    \vspace{1.2in}
    \begin{figure}[ht]
    \centering
   \includegraphics[trim=0 0 0 0, clip, scale = 0.4]{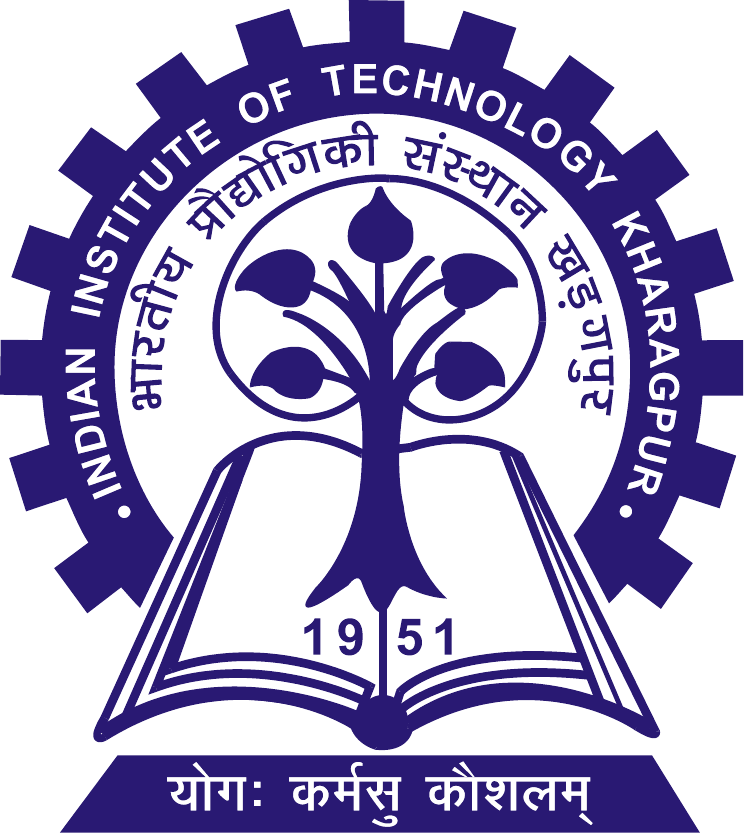}
    \end{figure}
    {\bf Electrical Engineering Department}\\
    {\bf Indian Institute of Technology, Kharagpur, India}\\
    {\bf  July 2019}
\end{center}
\end{titlepage}
\thispagestyle{empty}
\cleardoublepage
\thispagestyle{empty}
\setcounter{tocdepth}{4}
\tableofcontents
\cleardoublepage
\setcounter{page}{1}
\begin{abstract}
This is a collection of literature reviews and some initial work on the estimation of frequency of gravitational waves from a binary black hole merger for low SNR. This document provides a starting point, with a broad overview of the prerequisites required for the follow-up work on frequency estimation.
\end{abstract}
\section{Introduction}
Gravitational waves are the distortions/ripples in space time generated when massive objects move with extreme accelerations. They travel at the speed of light, carrying information about their source \cite{wiki:001,caltech}. Compact Binary Inspirals, one of the major sources of gravitational waves, are of the following types \cite{caltech}:
\begin{enumerate}
\item  Binary Neutron Star (BNS)
\item  Binary Black Hole (BBH)
\item  Neutron Star-Black Hole Binary (NSBH)
\end{enumerate}
\subsection{Binary Black Hole}
 A binary black hole (BBH) is a system consisting of two black holes closely orbiting each other. When such a pair of black holes merge, energy is dissipated in the form of gravitational waves, having distinctive waveforms that can be calculated using general relativity \cite{caltech,wiki:002}.
\subsubsection{Life Cycle of a Binary Black Hole}
The lifecycle of a binary black hole is defined as follows \cite{wiki:002}:

\textit{Inspiral:}The first stage is that of Inspiral which is simialar to a gradually shrinking orbit. The initial stages of this phase take a very long time, as the gravitational waves emitted are very weak due to the large distance between the black holes. In addition to the orbit getting smaller due to the emission of gravitational waves, there may be a loss of extra angular momentum due to interactions with other matter present, such as other stars. As the orbit shrinks, the orbiting speed increases, and gravitational wave emission increases. When the black holes are close the gravitational waves cause the orbit to shrink rapidly. The last stable orbit or innermost stable circular orbit (ISCO) is the innermost complete orbit before the transition from inspiral to~merger.

\textit{Merger:}This is followed by a plunging orbit in which the two black holes meet, followed by the merger. Gravitational wave emission peaks at this time. The binary system coalesces into a singular black hole.

\textit{Ringdown:}Immediately after the merger, the newly formed black hole will oscillate in shape between a distorted, elongated spheroid and a flattened spheroid. This ringing is damped in the next stage, called the ringdown, by the emission of gravitational waves. The distortions from the spherical shape rapidly reduce until the final stable sphere is present, with a possible slight distortion due to remaining spin.
\subsubsection{Dynamic modelling}
Post-Newtonian approximations~can be used for modelling the inspiral \cite{wiki:002}. These approximate the general relativity field equations adding extra terms to equations in Newtonian gravity. Orders used in these calculations may be termed 2PN (second order post Newtonian) 2.5PN or 3PN (third order post Newtonian). 

Effective-one-body (EOB) solves the dynamics of the binary black hole system by transforming the equations to those of a single object. This is especially useful where mass ratios are large, such as a stellar mass black hole~merging with a galactic core black hole, but can also be used for equal mass systems. 

For the ringdown, black hole perturbation theory can be used. The final Kerr black hole is distorted, and the spectrum of frequencies it produces can be calculated. To solve for the entire evolution, including merger, requires solving the full equations of general relativity. This can be done in~numerical relativity simulations. 

\subsection{LIGO}
The Laser Interferometer Gravitational-Wave Observatory (LIGO) is the one of the world's largest gravitational wave detectors \cite{wiki:003}. It accomplishes it's purpose of detecting gravitational waves with the help of laser interferometry. It exploits the properties of light and space to understand the sources of gravitational waves \cite{caltech}. But it cannot do this detection on its own, and more than one detectors are required to identify gravitational waves definitively \cite{caltech}. Other gravitational wave detectors include current ones like GEO600 \cite{wiki:004} \& VIRGO \cite{wiki:005}, and upcoming ones like LIGO India (INDIGO) \cite{wiki:006}, Einstein Telescope \cite{wiki:007} \& LISA \cite{wiki:008}.
\section{Literature Reviews}
\subsection{2016 - Observation of Gravitational Waves from a Binary Black Hole Merger \cite{caprini2016observation}}
Einstein predicted the existence of gravitational waves in 1916. This was after his final formulation of the general relativity field equations. He discovered that there existed wave solutions to linearized field equations. The waves were transverse and travelled at light speed carrying information about its source. 

On September 14, 2015 at 09:50:45 UTC, the LIGO Hanford, WA, and Livingston, LA, observatories detected the coincident signal GW150914. The signal sweeps upwards in frequency from 35 to 250 Hz with a peak gravitational-wave strain of $1.0 \times 10^{-21}$. It matches the waveform predicted by general relativity for the inspiral and merger of a pair of black holes and the ringdown of the resulting single black hole. The signal was observed with a matched-filter signal-to-noise ratio of 24. The source lies at a luminosity distance of 410 Mpc. The initial black hole masses are $36M _\odot$ and $29M _\odot$, and the final black hole mass is $62M _\odot$, with $3.0M _\odot$ radiated as gravitational waves. The basic features of GW150914 point to it being produced by the coalescence of two black holes—i.e., their orbital inspiral and merger, and subsequent final black hole ringdown. Over 0.2 s, the signal increases in frequency and amplitude in about 8 cycles from 35 to 150 Hz, where the amplitude reaches a maximum. The most plausible explanation for this evolution is the inspiral of two orbiting masses, $m_1$ and $m_2$, due to gravitational-wave emission. At the lower frequencies, such evolution is characterized by the chirp mass
\begin{align*}
{M_{ch}} = \frac{{{{({m_1}{m_2})}^{3/5}}}}{{{{({m_1} + {m_2})}^{1/5}}}} = \frac{{{c^3}}}{G}{\left[ {\frac{5}{{96}}{\pi ^{ - 8/3}}{f^{ - 11/3}}\dot f} \right]^{3/5}}
\end{align*}
where $f$ and $\dot{f}$ are the observed frequency and its time derivative and G and c are the gravitational constant and speed of light.
\subsection{2015 - Parameter estimation for compact binaries with ground-based gravitational wave observations using LALInference \cite{veitch2015parameter}}
The LALInference is a software containing a C library with several post-processing tools written in python. It uses the LSC Algorithm Library (LAL) to provide 
\begin{enumerate}
\item  Standard methods of accessing GW detector data; 
\item  all the waveform approximants included in LAL that describe the evolution of point mass binary systems for use; 
\item  Likelihood functions for the data observed by a network of ground-based laser interferometers given a waveform model and a set of model parameters; 
\item  Three independent stochastic sampling techniques of the parameter space to compute PDFs and evidence;
\item  Standard post-processing tools to generate probability credible regions for any set of parameters
\end{enumerate}
There are many models for the GW signal that may be emitted during a compact binary merger. These models are known as waveform families. Each waveform family can be thought of as a function that takes as input a parameter vector $\theta$ and produces as output $h_{+,\times } (t)$, either a time domain $h(\theta ;t)$ or frequency-domain $h(\theta ,f)$ signal. The parameter vector $\theta $ generally includes at least nine parameters: 
\begin{enumerate}
\item  Component masses $m_{1} $ and $m_{2} $. We use a reparameterization of the mass plane into the chirp mass,
\[M_{c} {}_{h} =(m_{1} m_{2} )^{3/5} (m_{1} +m_{2} )^{-1/5} \] 
%\end{enumerate}
and the asymmetric mass ratio 
\[q=m_{2} /m_{1} \] 
We take m1 $\mathrm{\ge}$ m2 when labelling the components. 
Another parametrisation is the symmetric mass ratio 
\[\eta =\frac{m_{1} m_{2} }{(m_{1} +m_{2} )^{2} } \] 
%\begin{enumerate}
\item  The luminosity distance to the source $d_{L} $; 
\item  The right ascension $\alpha $ and declination $\delta $ of the source; 
\item  The inclination angle $\iota $, between the system's orbital angular momentum and the line of sight. 
\item  The polarisation angle $\psi $ which describes the orientation of the projection of the binary's orbital momentum vector onto the plane on the sky,
\item  An arbitrary reference time $t_{c} $; 
\item  The orbital phase ${\phi _{c} }$ of the binary at the reference time $t_{c} $.
\end{enumerate}
Above parameters are necessary to describe a circular binary consisting of point-mass objects with no spins. If the spins of the binary's componenets are included, six additional parameters are required:
\begin{enumerate}
    \item dimensionless spin magnitude $a_i$, defined as ${a_i} \equiv \left| {{s_i}} \right|/m_i^2$ in range $[0, 1]$, where $s_i$ is the spin vector of the object $i$, and
    \item two angles for each $s_i$ specifying its orientation with respect to the plane defined by the line of sight of the initial orbital angular momentum.
\end{enumerate}
The time to coalescence of a binary emitting GWs at frequency $f$ is give as:
\begin{align*}
   \tau  = 93.9{\left( {\frac{f}{{30Hz}}} \right)^{ - 8/3}}{\left( {\frac{{{M_{ch}}}}{{0.87{M_ \odot }}}} \right)^{ - 5/3}}\sec
\end{align*}
The frequency of gravitational wave emission at the innermost stable circular orbit for a binary with non-spinning components is:
\[f_{isco} =\frac{1}{6^{3/2} \pi (m_{1} +m_{2} )} =4.4\left(\frac{M _\odot }{m_{1} +m_{2} } \right)kHz\] 
The low-frequency cut-off of the instrument, which sets the duration of the signal, was 40 Hz for LIGO in initial/enhanced configuration and 30 Hz for Virgo. The most commonly used waveform models:
\begin{enumerate}
\item  TaylorF2 \cite{buonanno2009comparison}
\item  SpinTaylorT4 \cite{buonanno2003detecting}
\item  IMRPhenomB \cite{ajith2011inspiral}
\item  EOBNRv2 \cite{pan2011inspiral}
\end{enumerate}
\subsection{1995 - Gravitational waves from inspiraling compact binaries: Parameter estimation using second-post-Newtonian waveforms \cite{poisson1995gravitational}}
Inspiraling compact binary parameters can be estimated using matched filtering of templates of gravitational waves to the actual observed output of gravitational wave detectors. Frequency sweeps are used to describe the time-dependent behaviour of a sample in the non-destructive deformation range.

Using a recently calculated formula, accurate to second post-Newtonian (2PN) order [order $(v/c)$ where $v$ is the orbital velocity], for the frequency sweep $(dF/dt)$ induced by gravitational radiation damping, the statistical errors in the determination of such source parameters as the "chirp mass" $M_{c} {}_{h} $, reduced mass $\mu $, and spin parameters $\beta $ and $\sigma $ (related to spin-orbit and spin-spin effects, respectively) can be studied. Previous results using template phasing accurate to 1.5PN order actually underestimated the errors in $M_{c} {}_{h} $, $\mu $, and $\beta $. Templates with 2PN phasing yield somewhat larger measurement error.

The gravitational-wave signal increases in amplitude as its frequency sweeps through the detector frequency bandwidth, from approximately 10Hz to 1000Hz (This characteristic is referred to as chirp). The signal measured by the detector is passed through a linear filter constructed from the expected signal $h(t;\theta )$ and the spectral density of the detector noise. The signal-to-noise ratio is then computed. The expected signal and the signal-to-noise ratio are expressed as functions of the vector $\theta $ which collectively represents the source parameters. The actual value of these parameters, which we denote $\tilde{\theta }$, is unknown prior to the measurement. When $\theta =\tilde{\theta }$ the linear filter becomes the Wiener optimum filter which is well known to yield the largest possible signal to noise ratio. The source parameters can therefore be determined by maximizing SNR over a broad collection of expected signals $h(t;\theta )$, loosely referred to as "templates.

The gravitational-wave signal from an inspiraling compact binary can be calculated exactly using general relativity (this would require the numerical integration of Einstein's equations). In practice, however, some approximation scheme like post Newtonian and post Minkowskian expansion is required. The binary inspiral waveform as measured by the detector can be modelled as
\[h(t;\theta )=r^{-1} Q(angles)M_{ch} (\pi M_{ch} F)^{2/3} \cos \Phi (t)\] 
Here, $r$ is the distance to the source, $Q$ is a function of various angles, $F(t)$ is the gravitational wave frequency \& $\Phi (t)=\int 2\pi F(t)dt $ the phase. Everything is in geometrized units i.e $G=c=1$. The chirp mass is represented as
\[M_{ch} =\eta ^{3/5} M\] 
Where the total mass
\[M=m_{1} +m_{2} \] 
And symmetric mass ratio
\[\eta =\mu /M\] 
Where reduced mass
\[\mu =m_{1} m_{2} /(m_{1} +m_{2} )\] 
Now the 2PN expansion of the frequency sweep is:
\begin{align*}
\frac{{dF}}{{dt}} = \frac{{96}}{{5\pi {M_{ch}}}}{(\pi {M_{ch}}F)^{11/3}}\left[ {1 - \left( {\frac{{743}}{{336}} + \frac{{11}}{4}\eta } \right){{(\pi MF)}^{2/3}}} \right.\\
 + (4\pi  - \beta )(\pi MF)\\
\left. { + \left( {\frac{{34103}}{{18144}} + \frac{{13661}}{{2016}}\eta  + \frac{{59}}{{18}}{\eta ^2} + \sigma } \right){{(\pi MF)}^{4/3}}} \right]
\end{align*} 
Where $\beta$ and $\sigma$ are the spin-orbit and spin-spin parameters respectively, given by
\begin{align*}
    \begin{gathered}
  \beta  = \frac{1}{{12}}\sum\limits_{i = 1}^2 {\left[ {113{{\left( {{m_i}/M} \right)}^2} + 75\eta } \right]\hat L.{\chi _i}}  \hfill \\
  \sigma  = \frac{\eta }{{48}}\left( { - 247{\chi _1}.{\chi _2} + 721\hat L.{\chi _1}\hat L.{\chi _2}} \right) \hfill \\ 
\end{gathered}
\end{align*}
where $\chi _i = s_i/m_i^2$. The authors have taken both spin parameters as zero in their calculations and for a binary black hole system the masses have been considered as 10 solar masses for both the black holes. They have also considered 10Hz as the frequency entering the bandwidth of the detector. The frequency sweep equation is integrated to obtain the phase and time as functions of the gravitational wave frequency:
\begin{align*}
\Phi (F) = {\phi _c} - \frac{1}{{16}}{(\pi {M_{ch}})^{ - 5/3}}\left[ {1 + \frac{5}{3}\left( {\frac{{743}}{{336}} + \frac{{11}}{4}\eta } \right){{(\pi MF)}^{2/3}}} \right.\\
 - \frac{5}{2}(4\pi  - \beta )(\pi MF)\\
\left. { + 5\left( {\frac{{3058673}}{{1016064}} + \frac{{5429}}{{1008}}\eta  + \frac{{617}}{{144}}{\eta ^2} - \sigma } \right){{(\pi MF)}^{4/3}}} \right]
\end{align*}
\begin{align*}
t(F) = {t_c} - \frac{5}{{256}}{M_{ch}}{(\pi {M_{ch}})^{ - 8/3}}\left[ {1 + \frac{4}{3}\left( {\frac{{743}}{{336}} + \frac{{11}}{4}\eta } \right){{(\pi MF)}^{2/3}}} \right.\\
 - \frac{8}{5}(4\pi  - \beta )(\pi MF)\\
\left. { + 2\left( {\frac{{3058673}}{{1016064}} + \frac{{5429}}{{1008}}\eta  + \frac{{617}}{{144}}{\eta ^2} - \sigma } \right){{(\pi MF)}^{4/3}}} \right]
\end{align*}
This signal cannot be allowed to reach arbitrarily large frequency values; so, the cut off ${F_{i}}$ corresponding to the end of the inspiral is set according to the relation:
\[F_{i} =\frac{6^{-3/2} }{\pi M} \] 
Which is same as the frequency at the innermost stable circular orbit (ISCO) and comes from
\[\pi MF_{i} =(M/r_{i} )^{2/3} =6^{-3/2} \] 
Where
\[r_{i} =6M\] 
Is the Swarzschild radius of the ISCO for a test mass moving in the gravitational field of a mass $M$.
\subsection{1996 - Gravitational Waves from coalescing binaries: Detection strategies and Monte Carlo estimation of parameters \cite{balasubramanian1996gravitational,balasubramanian1996erratum}}
Strain at the detector is given by
\[h(t)=A[\pi f(t)]^{2/3} \cos [\phi (t)]\] 
Here f(t) is the instantaneous gravitational wave frequency, A is a constant that contains information about the distance from detector to the binary, the reduced and total mass of the system, and the antenna pattern of the detector. The phase can be expanded into components signifying the different post-Newtonian contributions as
\[\phi (t)=\phi _{0} (t)+\phi _{1} (t)+\phi _{1.5} (t)+...\] 
where $\phi _{0} (t)$ is the dominant Newtonian part of the phase and $\phi _{n} $ is the nth order correction to it. The Newtonian part is the only one included in quadruple approximation meaning
\[\phi (t)=\phi _{0} (t)=\frac{16\pi f_{a} \tau _{0} }{5} \left(1-\left(\frac{f}{f_{a} } \right)^{-5/3} \right)+\Phi \] 
Where f(t) is the instantaneous frequency of the gravitational wave given by
\[t-t_{a} =\tau _{0} \left[1-\left(\frac{f}{f_{a} } \right)^{-8/3} \right]\] 
$\tau _{0} $ is a constant having dimensions of time given as
\[\tau _{0} =\frac{5}{256} M_{c} {}_{h} {}^{-5/3} (\pi f_{a} )^{-8/3} \] 
And $f_{a} $  \& $\Phi $ are the frequency and phase at $t=t_{a} $ . For correction up to 2PN order
\[\phi (t)=\phi _{0} (t)+\phi _{1} (t)+\phi _{1.5} (t)+\phi _{2} (t)\] 
Where
\[\phi _{1} (t)=4\pi f_{a} \tau _{1} \left[1-\left(\frac{f}{f_{a} } \right)^{-1} \right]\] 
\[\phi _{1.5} (t)=-5\pi f_{a} \tau _{1.5} \left[1-\left(\frac{f}{f_{a} } \right)^{-2/3} \right]\] 
\[\phi _{2} (t)=8\pi f_{a} \tau _{2} \left[1-\left(\frac{f}{f_{a} } \right)^{-1/3} \right]\] 
Now the instantaneous gravitational wave frequency of order 2PN is given by
\[\begin{array}{rcl} {t-t{}_{a} } & {=} & {\tau _{0} \left[1-\left(\frac{f}{f_{a} } \right)^{-8/3} \right]+\tau _{1} \left[1-\left(\frac{f}{f_{a} } \right)^{-2} \right]} \\ {} & {} & {+\tau _{1.5} \left[1-\left(\frac{f}{f_{a} } \right)^{-5/3} \right]+\tau _{2} \left[1-\left(\frac{f}{f_{a} } \right)^{-4/3} \right]} \end{array}\] 
The $\tau $'s are constant in the above equations, having dimensions of time which are dependent only on the masses of the binary and the frequency cut-off of the detector $f_{a} $. Now the chirp time has four components in total
\[\tau _{0} =\frac{5}{256} M_{c} {}_{h} {}^{-5/3} (\pi f_{a} )^{-8/3} \] 
\[\tau _{1} =\frac{5}{192\mu (\pi f_{a} )^{2} } \left(\frac{743}{336} +\frac{11}{4} \eta \right)\] 
\[\tau _{1.5} =\frac{1}{8\mu } \left(\frac{M}{\pi ^{2} f_{a}^{5} } \right)^{1/3} \] 
\[\tau _{2} =\frac{5}{128\mu } \left(\frac{M}{\pi ^{2} f_{a} {}^{2} } \right)^{2/3} \left(\frac{3058673}{1016004} +\frac{5429}{1008} \eta +\frac{617}{144} \eta ^{2} \right)\] 
Now, $t_{c} $ is the sum of arrival time and total chirp time \& $\Phi _{c} $ is a combination of $\Phi $ and various chirp times
\[t_{c} =t_{a} +\tau _{0} +\tau _{1} -\tau _{1.5} +\tau _{2} \] 
\[\Phi _{c} =\Phi +\frac{16\pi f_{a} }{5} \tau _{0} +4\pi f_{a} \tau _{1} -5\pi f_{a} \tau _{1.5} +8\pi f_{a} \tau _{2} \] 
The Fourier transform of the 2PN waveform is as follows
\[\tilde{h}(f)={\rm N} f^{-7/6} \exp \left[i\sum _{u=1}^{6}\phi _{u} (f)\lambda ^{u} -i\frac{\pi }{4}  \right]\] 
Where
\[{\rm N} =A\pi ^{2/3} \left(\frac{2\tau _{0} }{3} \right)^{1/2} f_{a} {}^{4/3} \] 
is a normalization constant, $\lambda ^{u} $ , $u=1,....,6$ represents the various post Newtonian parameters
\[\lambda ^{u} =\{ t_{a} ,\Phi ,\tau _{0} ,\tau _{1} ,\tau _{1.5} ,\tau _{2} \} \] 
And
\[\phi _{1} =2\pi f\] 
\[\phi _{2} =-1\] 
\[\phi _{3} =2\pi f-\frac{16\pi f_{a} }{5} +\frac{6\pi f_{a} }{5} \left(\frac{f}{f_{a} } \right)^{-5/3} \] 
\[\phi _{4} =2\pi f-4\pi f_{a} +2\pi f_{a} \left(\frac{f}{f_{a} } \right)^{-1} \] 
\[\phi _{5} =-2\pi f+5\pi f_{a} -3\pi f_{a} \left(\frac{f}{f_{a} } \right)^{-2/3} \] 
\[\phi _{6} =2\pi f-8\pi f_{a} +6\pi f_{a} \left(\frac{f}{f_{a} } \right)^{-1/3} \] 
\subsection{1998 - Estimation of parameters of gravitational wave signals from coalescing binaries \cite{balasubramanian1998estimation}}
The strain $s(t)$ at the detector is given by
\begin{align*}
    s(t) = A{(\pi f(t))^{2/3}}\cos (\phi (t) = \Phi )
\end{align*}
here $\Phi$ is the initial phase of the wave at some fiducial time $t=t_s$. The phase of the waveform $\phi(t)$ contains several pieces corresponding to different post-Newtonian contributions which can be schematically written as
\begin{align*}
    \phi (t) = {\phi _0}(t) + {\phi _1}(t) + {\phi _{1.5}}(t) + ...
\end{align*}
\subsection{1998 - Bayesian Bounds on parameter estimation accuracy for compact binary gravitational wave signals \cite{nicholson1998bayesian}}
The coalescing binary inspiral waveform can be represented as
\[h(t)=A(\pi f(t))^{2/3} \cos (\Phi (t))\] 
Frequency and phase read
\[f(t)=f_{a} \left(1-\frac{t-t_{a} }{\tau } \right)^{8/3} \] 
And
\[\Phi (t)=\frac{16\pi f_{a} \tau }{5} \left(1-\left(\frac{f(t)}{f_{a} } \right)^{-5/3} \right)+\Phi _{a} \] 
Where the chirp time $\tau $ can be cast in terms of chirp mass as
\[\tau =\frac{5}{256} M_{ch} (\pi f_{a} )^{-8/3} \] 
The frequency and phase of the signal is $f_{a} $ \& $\Phi _{a} $ respectively at the time $t=t_{a} $. The waveform is characterized in terms of 3 parameters $t_{a} $,$\Phi _{a} $ \& $\tau $. The signal's Fourier transform $\tilde{h}(f)$ in the stationary phase approximation reads
\[\tilde{h}(f)={\rm N} f^{-7/6} \exp (\Psi (f))\] 
Where
\[{\rm N} =A\pi ^{2/3} \left(\frac{2\tau }{3} \right)^{1/2} f_{a} {}^{4/3} \] 
Is a normalisation constant, and
\[\Psi (f)=i\sum _{v=1}^{3}\psi _{v} \lambda ^{v} -i\frac{\pi }{4}  \] 
$\lambda ^{v} $ represents the parameter vector
\[\lambda ^{v} \equiv (t_{a} ,\Phi _{a} ,\tau )\] 
And
\[\begin{array}{l} {\psi _{1} =2\pi f} \\ {\psi _{2} =-1} \\ {\psi _{3} =2\pi f-\frac{16\pi f_{a} }{5} +\frac{6\pi f_{a} }{5} \left(\frac{f}{f_{a} } \right)^{-5/3} } \end{array}\]
\subsection{2005 - Parameter Estimation of inspiraling compact binaries using 3.5 post Newtonian gravitational wave phasing: The non-spinning case \cite{arun2005parameter}}
The problem is treated perturbatively by expanding the general relativistic equations of motion and wave generationas a power series in $v/c$, where $v$ is the characteristic orbital velocity of the system. The phase is computed at the highest PN order available, but the amplitude is taken to be Newtonian. The best template is one with phasing at 3.5PN and amplitude at 2.5PN. Parameter estimation worsens ten-fold due to inclusion of $\betaup$. Comparatively the effect of $\sigmaup$ is less drastic. It worsens estimation only by a factor of unity. Also, spin effects beyond 2PN have not yet been computed. The output of the detector includes both signal and noise represented as
\[x(t)=h(t)+n(t)\] 
where $x(t)$ is the signal registered and $n(t)$ is the noise, assumed to be Gaussian, with a zero mean. In going from lower to higher post Newtonian orders it is observed that there is an oscillation of errors in the chiro mass and reduced mass. 3.5PN always gives smaller errors than 2PN. The frequency sweep provides a way of partially computing the dependence of the wave amplitude on different PN orders. The expansion $\dot{F}$ up to 3.5PN is given by
\begin{align*}
\frac{{dF}}{{dt}} = \frac{{96}}{{5\pi {M_{ch}}}}{(\pi {M_{ch}}F)^{11/3}}\left[ {1 - \left( {\frac{{743}}{{336}} + \frac{{11}}{4}\eta } \right){{(\pi MF)}^{2/3}}} \right.\\
 + (4\pi )(\pi MF)\\
 + \left( {\frac{{34103}}{{18144}} + \frac{{13661}}{{2016}}\eta  + \frac{{59}}{{18}}{\eta ^2}} \right){(\pi MF)^{4/3}}\\
 + \left( { - \frac{{4159\pi }}{{672}} - \frac{{173\pi }}{8}\eta } \right){(\pi MF)^{5/3}}\\
\left[ {\frac{{16447322263}}{{139708800}} + \frac{{16{\pi ^2}}}{3} - \frac{{1712}}{{105}}\gamma } \right.\\
 + \left( { - \frac{{273811877}}{{1088640}} + \frac{{451{\pi ^2}}}{{48}} - \frac{{88}}{3}\theta  + \frac{{616}}{9}\lambda } \right)\eta \\
\left. { + \frac{{541}}{{896}}{\eta ^2} - \frac{{5605}}{{2592}}{\eta ^3} - \frac{{856}}{{106}}\log (16x)} \right]{(\pi MF)^2}\\
 + \left. {\left( { - \frac{{4415}}{{4032}} + \frac{{661775}}{{12096}}\eta  + \frac{{149789}}{{3024}}{\eta ^2}} \right)\pi {{(\pi MF)}^{7/3}}} \right]
\end{align*}
Where the Euler's constant $\gamma =0.577...$ and the coefficients $\lambda =-\frac{1987}{3080} \simeq -0.6451$,$\theta =-\frac{11831}{9240} \simeq -1.28$. There is an erratum in the above formula. The variable $x$ which has not been defined, is the guage-invariant later shown in this document.
\subsection{2004 - Gravitational Waves from Mergin Compact Binaries: How Accurately Can One Extract the Binary's Parameters from the Inspiral Waveform? \cite{cutler1994gravitational}}
When using geometrized units $c=G=1$, all quantities are measured in units of seconds, except where solar masss is required. The conversion factor used is $1M _\odot = 4.926 \times 10^{-6} sec$. This is calculated as ${M _\odot \times G} / c^3$ where $M _\odot$ is the solar mass in kg with $c$ and $G$ representing speed of light and gravitaional constant respectively in SI units.
\subsection{2000 - Frequency-domain P-approximant filter for time-truncated inspiral gravitational wave signals for compact binaries \cite{damour2000frequency}}
The PN approximation is basically a Taylor expansion (in powers of $v/c$). Templates based on such straightforward PN expansions may be termed as ``Taylor approximants''. These expansions have a very slow convergence and have an oscillatory behaviour. Post-Newtonian theory is an approximate version of general relativity that applies when the gravitational field is weak, and the motion of the matter is slow. A correction of order $(v/c)^{n}$ to a Newtonian expression is said to be of $(n/2)$ PN order.
\subsection{2017 - Parameter estimation method that directly compares gravitational wave observations to numerical relativity \cite{lange2017parameter}}
Numerical Relativity is now used to simulate the late inspiral, merger and final ringdown of black hole binary systems. The binary is parametrized using the mass ratio $q=m_1/m_2$ where $m_1 \geq m_2$ and dimensionless spin paramters as ${\chi _i} = {S_i}/{m_i}^2$ where $i=1,2$ are the indexes to the two black holes. The effective spin is ${\chi _{eff}}=({S_1}/{m_1} + {S_2}/{m_2}).\hat L/M$. The authors have used a total NR template bank of 1002 simulations which includes 313 from Simulating Extreme Spacetimes (SXS) group, 407 from Rochesteer Institute of Technology (RIT) group and 282 from Georgia Tech (GT) group. They then use likelihood \& KL divergence to try to estimate parameters by comparing the observation with the template bank.
\subsection{2019 - Analytical Black-Hole Binary Merger Waveforms \cite{mcwilliams2019analytical}}
The authors present a highly accurate, fully analytical model for the late inspiral, merger, and ringdown of black-hole binaries with arbitrary mass ratios and spin vectors and the associated gravitational radiation, including the contributions of harmonics beyond the fundamental mode. Their model assumes only that nonlinear eﬀects remain small throughout the entire coalescence, and is developed based on a physical understanding of the dynamics of late stage binary evolution. The "merger-ringdown" is the part of the waveform occurring after the peak amplitude has been reached. The "late inspiral" is the part occuring just after the system reaches the innermost stable circular orbit (ISCO) of the final black hole but before the light ring. 

The authors refer to their approach as the Backwards One-Body (BOB) method. The gravitational-wave emission of a single perturbed black hole is well described by the properties of null geodesics on unstable circular orbits at the black hole’s light ring. They also talk about an anomaly, which is the dichotomy in the waveform at merger and also at a time long after the merger both correspond to disturbances at the light ring due to different assumptions. Null geodesics are the geodesics of light as they are particles with null masses. A geodesic is the path of an object in curved space-time. An example of this is the path of the object free from all external gravitational forces \cite{wiki:009}. These geodesics are described by a set of coordinates ${t, r, \theta, \phi}$, and their evolution is expressed as
\begin{align*}
\begin{array}{l}
t = {t_p} + \eta  + \epsilon\mathfrak{h}(t - {t_p})\\
r = {r_{1r}}\left[ {1 + \epsilon\mathfrak{f}(t - {t_p})} \right]\\
\theta  = \frac{\pi }{2}\left[ {1 + \epsilon\mathfrak{p}(t - {t_p})} \right]\\
\phi  = \omega \left[ {t + (t - \epsilon\mathfrak{g}{t_p})} \right]
\end{array}
\end{align*}
where $t_p$ is the time when the congruence converges, corresponding to the peak waveform amplitude, $\eta$ is an affine parameter, $\epsilon$ is a small dimensionless order-counting parameter, $r_{lr}$ is the light-ring radius, $\omega$ is the orbital frequency of the geodesic, and $\mathfrak{f}$, $\mathfrak{g}$, $\mathfrak{h}$, and $\mathfrak{p}$ are functions determined from the requirements that the perturbed orbits are still null geodesics, and that $\mathfrak{f}(0) = \mathfrak{g}(0) = \mathfrak{h}(0) = \mathfrak{p}(0) = 0$. These perturbation functions are given by
\begin{align*}
\begin{gathered}
  \mathfrak{f} = \sinh \left[ {\gamma \left( {t - {t_p}} \right)} \right] \hfill \\
  \mathfrak{g} = 0 \hfill \\
  \mathfrak{h} = 2\frac{\omega }{{{\gamma ^2}}}\sqrt {\frac{{3M}}{{{r_{1r}}}}} \left\{ {1 - \cosh \left[ {\gamma \left( {t - {t_p}} \right)} \right]} \right\} \hfill \\
  \mathfrak{p} = 0 \hfill \\ 
\end{gathered}
\end{align*}
where $\gamma$ is the Lyapunov exponent of the congruence, and corresponds in the wave picture to the inverse damping time of the amplitude. The waveform amplitude is represented as
\begin{align*}
    A = A_p sech [ \gamma ( t - t_p ) ]
\end{align*}
The authors will present the full details of their calculations in a followup work, summarizing that, they solve an approximation to Zerilli equation which describes the scattering of gravitational perturbations by a black hole to first order in the black-hole spin. The dominant contribution to the gravitational-wave emission comes from gravitational perturbations scattering off of the curvature potential rather than arriving directly from the effective perturber, which are described by the Zerilli equation. Now, for quasicircular systems $|h| \approx |\psi _4| / \omega ^2 $, where
\begin{align*}
    |\psi _4| = A_p sech [ \gamma (t - t_p)]
\end{align*}
The above equation can be combined with an analytical model for $\omega$ to define an analytical model for strain amplitude. 
\subsection{2017 - Complete waveform model for compact binaries on eccentric orbits \cite{huerta2017complete}}
The authors present a time domain waveform model that describes the inspiral, merger and ringdown of compact binary systems whose components are spin less, and orbits having low to moderate eccentricity. An enhanced prescription for the inspiral evolution, based on post-Newtonian approximations, is combined with a fully analytical prescription for the merger-ringdown evolution constructed using a catalog of numerical relativity. This model can be immediately used in the context of aLIGO to
\begin{enumerate}
    \item quantify the sensitivity of quasicircular searches and burst searches to eccentric signals, 
    \item study template bank construction for nonspinning, eccentric BBHs,
    \item estimate the eccentricity of detected BBH signals, under the assumption that the binary components are not spinning, and 
    \item explore the sensitivity of burst-like searches that have been tuned to detect highly eccentric systems (e0 ~1) to recover signals with moderate values of eccentricity.
\end{enumerate}
Key features of the model are: 
\begin{enumerate}
 \item It includes third-order PN accurate expansions for eccentric orbits both for the equations of motion of the binary and its far-zone radiation field. The radiative evolution includes instantaneous, tails and tails-of-tails contributions, and a contribution due to nonlinear memory. 
 \item The accuracy of the inspiral evolution is improved by including 3.5PN corrections for quasicircular orbits (at all powers of the symmetric mass ratio). 
 \item To further improve phase accuracy especially for unequal-mass systems, the 3PN accurate inspiral evolution for eccentric systems is corrected by including up to 6PN terms both for the energy flux of quasicircular binaries and gravitational self-force corrections to the binding energy of compact binaries at first order in the symmetric mass ratio $\eta$. 
 \item We combine the aforementioned enhanced inspiral evolution with a merger and ringdown treatment using the implicit rotating source (IRS) formalism, fitted against NR simulations up to mass ratio 10.
\end{enumerate}
The equations of motion are parametrized in terms of the mean orbital frequency $\omega$ through the gauge-invariant quantity $x = (M \omega)^{2/3}$, and the temporal eccentricity $e_t \equiv e$. In the context of eccentric binaries $\omega = \langle \dot \phi \rangle = Kn$,where the average $\langle \rangle$ is taken over an orbital period. The mean motion $n$ is related to the mean anomaly $l$ through the relation $M \dot l = M n$, $\dot \phi$ is the instantaneous angular velocity, and the periastron precession $K$ and relativistic precession $k$ are related through $K = 1+k$. At 3PN order, the Keplerian parametrization of the orbit in terms of the magnitude of the relative separation vector $r$, and the mean anomaly $l$ is given by
\begin{align*}
\begin{gathered}
  \frac{r}{M} = \frac{{1 - e\cos u}}{x} + \sum\limits_{i = 1}^{i = 3} {{r_{iPN}}{x^{i - 1}}}  \hfill \\
  l = u - e\sin u + \sum\limits_{i = 2}^{i = 3} {{l_{iPN}}{x^i}}  \hfill \\ 
\end{gathered} 
\end{align*}
The orbital evolution has two components. The conservative piece is derived from a PN Hamiltonian including corrections at 3PN order and has the form
\begin{align*}
\begin{gathered}
  M\dot \phi  = {{\dot \phi }_{0PN}}{x^{3/2}} + {{\dot \phi }_{1PN}}{x^{5/2}} + {{\dot \phi }_{2PN}}{x^{7/2}} + {{\dot \phi }_{3PN}}{x^{9/2}} \hfill \\
  M\dot l = Mn = n{x^{3/2}} + {n_{1PN}}{x^{5/2}} + {n_{2PN}}{x^{7/2}} + {n_{3PN}}{x^{9/2}} \hfill \\ 
\end{gathered}
\end{align*}
The gauge-invariant expansion parameter $x$ and the eccentricity $e$ evolve as:
\begin{align*}
    \begin{gathered}
  M\dot x = {{\dot x}_{0PN}}{x^5} + {{\dot x}_{1PN}}{x^6} + {{\dot x}_{2PN}}{x^7} + {{\dot x}_{3PN}}{x^8} + {{\dot x}_{HT}} \hfill \\
  M\dot e = {{\dot e}_{0PN}}{x^4} + {{\dot e}_{1PN}}{x^5} + {{\dot e}_{2PN}}{x^6} + {{\dot e}_{3PN}}{x^7} + {{\dot e}_{HT}} \hfill \\ 
\end{gathered}
\end{align*}
The PN waveform strain is constructed as follows
\begin{align*}
    {h^{inspiral}}(t) = {h_ + }^{inspiral}(t) - i{h_ \times }^{inspiral}(t)
\end{align*}
with the plus and cross polarizations given by
\begin{align*}
\begin{gathered}
  {h_ + } =  - \frac{{M\eta }}{R}\left\{ {\left( {{{\cos }^2}\iota  + 1} \right)\left[ {\left( { - {{\dot r}^2} + {r^2}{{\dot \phi }^2} + \frac{M}{r}} \right)\cos 2\Phi  + 2r\dot r\dot \phi \sin 2\Phi } \right] + \left( { - {{\dot r}^2} - {r^2}{{\dot \phi }^2} + \frac{M}{r}} \right){{\sin }^2}\iota } \right\} \hfill \\
  {h_ \times } =  - \frac{{2M\eta }}{R}\cos \iota \left\{ {\left( { - {{\dot r}^2} + {r^2}{{\dot \phi }^2} + \frac{M}{r}} \right)\sin 2\Phi  - 2r\dot r\dot \phi \cos 2\Phi } \right\} \hfill \\ 
\end{gathered}
\end{align*}
where $\Phi = \phi - \chi ,(\chi,\iota)$ represent the polar angles of the observer $R$ is the distance to the binary. The innermost stable circular orbit (ISCO) frequency is given by
\begin{align*}
    {f_{ISCO}} = \frac{1}{{\pi M}}{\left( {\frac{{1 + e}}{{6 + 2e}}} \right)^{3/2}}
\end{align*}
The number of GW cycles $N$ is defined as
\begin{align*}
    N = \frac{1}{\pi }\left[ {\left\langle \phi  \right\rangle \left( {{f_{ISCO}}} \right) - \left\langle \phi  \right\rangle \left( {{f_{\min }}} \right)} \right]
\end{align*}
where $f_{min} = 15Hz$. Now the time-derivative of the gauge-invariant for the 3PN model with eccentrity taken as zero is represented as
\begin{align*}
\begin{gathered}
  M{\left. {\frac{{dx}}{{dt}}} \right|_{e \to 0}} = \frac{{64}}{5}\eta {x^5}\left\{ {1 + \left( { - \frac{{743}}{{336}} - \frac{{11}}{4}\eta } \right)x + 4\pi {x^{3/2}} + \left( {\frac{{34103}}{{18144}} + \frac{{13661}}{{2016}}\eta  + \frac{{59}}{{18}}{\eta ^2}} \right){x^2}} \right. \hfill \\
   + \left( { - \frac{{4159\pi }}{{672}} - \frac{{189\pi }}{8}\eta } \right){x^{5/2}} + \left[ {\frac{{16447322263}}{{139708800}} - \frac{{1712\gamma }}{{105}} + \frac{{16{\pi ^2}}}{3} - \frac{{856}}{{105}}\log \left( {16x} \right)} \right. \hfill \\
  \left. {\left. { + \left( { - \frac{{56198689}}{{217728}} + \frac{{451{\pi ^2}}}{{48}}} \right)\eta  + \frac{{541}}{{896}}{\eta ^2} - \frac{{5605}}{{2592}}{\eta ^3}} \right]{x^3}} \right\} \hfill \\ 
\end{gathered}
\end{align*}
where $\gamma$ is Euler's constant. Also the time evolution of eccentricity $e$, phase $\phi$, and mean anomaly $l$ is given by
\begin{align*}
    \begin{gathered}
  {\left. {M\frac{{d\phi }}{{dt}}} \right|_{e \to 0}} = {x^{3/2}} \hfill \\
  {\left. {M\frac{{de}}{{dt}}} \right|_{e \to 0}} = 0 \hfill \\
  {\left. {M\frac{{dl}}{{dt}}} \right|_{e \to 0}} = {x^{3/2}}\left\{ {1 + 3x + \left( {7\eta  - \frac{9}{2}} \right){x^2} + \left( { - \frac{{27}}{2} + \left( {\frac{{481}}{4} - \frac{{123}}{{32}}{\pi ^2}} \right)\eta  - 7{\eta ^2}} \right){x^3}} \right\} \hfill \\ 
\end{gathered}
\end{align*}
After including 3.5PN corrections we get
\begin{align*}
    {\left. {M\frac{{dx}}{{dt}}} \right|_{e \to 0\{ 3.5PN\} }} = {\left. {M\frac{{dx}}{{dt}}} \right|_{e \to 0\{ 3PN\} }} + \frac{{64\pi }}{5}\eta {x^5}\left[ { - \frac{{4415}}{{4032}} + \frac{{3658675}}{{6048}}\eta  + \frac{{91945}}{{1512}}{\eta ^2}} \right]{x^{7/2}}
\end{align*}
Now using 6PN correction to energy flux
\begin{align*}
    {\left. {M\frac{{dx}}{{dt}}} \right|_{e \to 0\{ 6PN\} }} = {\left. {M\frac{{dx}}{{dt}}} \right|_{e \to 0\{ 3.5PN\} }} + \frac{{64\pi }}{5}\eta {x^5}\left[ {{a_4}{x^4} + {a_{9/2}}{x^{9/2}} + {a_5}{x^5} + {a_{11/2}}{x^{11/2}} + {a_6}{x^6}} \right]
\end{align*}
where, the values of all constants of form $a_i$ can be found in the original literature's appendix. Now, for the merger-ringdown, the frequency evolution $\omega(t)$ and waveform amplitude $A(t)$ is given by
\begin{align*}
    \begin{gathered}
  \omega (t) = {\omega _{QNM}}(1 - \hat f) \hfill \\
  {\omega _{QNM}} = 1 - 0.63{(1 - {{\hat s}_{fin}})^{0.3}} \hfill \\
  A(t) = \frac{{{A_0}}}{{\omega (t)}}\left[ {\frac{{\left| {\dot \hat f} \right|}}{{1 + \alpha ({{\hat f}^2} - {{\hat f}^4})}}} \right] \hfill \\ 
\end{gathered}
\end{align*}
where ${{\hat s}_{fin}}$ is the spin of the black hole remnant. Also
\begin{align*}
    \begin{gathered}
  \hat f = \frac{c}{2}{\left( {1 + \frac{1}{\kappa }} \right)^{1 + \kappa }}\left[ {1 - {{\left( {1 + \frac{1}{\kappa }{e^{ - 2t/b}}} \right)}^{ - \kappa }}} \right] \hfill \\
  {{\hat s}_{fin}} = 2\sqrt 3 \eta  - \frac{{390}}{{79}}{\eta ^2} + \frac{{2379}}{{287}}{\eta ^3} - \frac{{4621}}{{276}}{\eta ^4} \hfill \\ 
\end{gathered}
\end{align*}
and the constants
\begin{align*}
    \begin{gathered}
  b(\eta ) = \frac{{16014}}{{979}} - \frac{{29132}}{{1343}}{\eta ^2} \hfill \\
  c(\eta ) = \frac{{206}}{{903}} + \frac{{180}}{{1141}}\sqrt \eta   + \frac{{424}}{{1205}}\frac{{{\eta ^2}}}{{\log (\eta )}} \hfill \\
  \kappa (\eta ) = \frac{{713}}{{1056}} - \frac{{23}}{{193}}\eta  \hfill \\
  \alpha (\eta ) = \frac{1}{{{Q^2}({{\hat s}_{fin}})}}\left( {\frac{{16313}}{{562}} + \frac{{21345}}{{124}}\eta } \right) \hfill \\
  Q({{\hat s}_{fin}}) = \frac{2}{{{{(1 - {{\hat s}_{fin}})}^{0.45}}}} \hfill \\ 
\end{gathered}
\end{align*}
The merger waveform can be obtained by
\begin{align*}
    \begin{gathered}
  {h^{merger}}(t) = {h_ + }^{merger}(t) - i{h_ \times }^{merger}(t) = A(t){e^{ - {\Phi _{gIRS}}(t)}} \hfill \\
  {\Phi _{gIRS}}(t) = \int_{{t_0}}^t {\omega (t)dt}  \hfill \\ 
\end{gathered}
\end{align*}
\subsection{2009 - Comparision of post-Newtonian templates for compact binary inspiral signals in gravitational-wave detectors \cite{buonanno2009comparison}}
The evolution of the orbital phase and the emitted gravitational radiation are now known to a rather high order up to $\mathcal{O} ({v^8})$, $v$ being the characteristic velocity of the binary. But due to an inherent freedom in choice of parameter used in PN expansions, and also the dependency on the method for solving the differential equations, the orbital evolution cannot be uniquely specified. The authors main aim is to determine the differences and similarities between different PN waveform families. They conclude that as long as the total mass remains less than a certain upper limit $M_{crit}$, all template families at 3.5PN order (except TaylorT3 and TaylorEt) are equally good for the purpose of detection. The value of $M_{crit}$ is found to be $~ 12 M _\odot$ for Initial, Enhanced and Advanced LIGO. From a purely computational point of view we recommend that 3.5PN TaylorF2 be used below $M_{crit}$ and EOB calibrated to numerical relativity simulations be used for total binary mass $M > M_{crit}$.

Post-Newtonian approximation computes the evolution of the orbital phase $\phi(t)$ of a compact binary as a perturbative expansion in a small parameter, typically taken as $v = (\pi MF)^{1/3}$ (characteristic velocity in the binary), or $x = v^2$, although other variants exist. Here $M$ is the total mass of the binary and $F$ the gravitational-wave frequency. Out of all the various models explained in the paper, the TaylorT3 has a time dependent frequency equation which is given as
\begin{align*}
\begin{gathered}
  F_{3.5}^{(T3)}(t) = \frac{{{\theta ^3}}}{{8\pi M}}\left[ {1 + \left( {\frac{{743}}{{2688}} + \frac{{11}}{{32}}\eta } \right){\theta ^2} - \frac{{3\pi }}{{10}}{\theta ^3} + \left( {\frac{{1855099}}{{14450688}} + \frac{{56975}}{{258048}}\eta  + \frac{{371}}{{2048}}{\eta ^2}} \right){\theta ^4}} \right. \\ 
   - \left( {\frac{{7729}}{{21504}} - \frac{{13}}{{256}}\eta } \right)\pi {\theta ^5} \\ 
   + \left\{ {\frac{{720817631400877}}{{288412611379200}}} \right. + \frac{{53}}{{200}}{\pi ^2} + \frac{{107}}{{280}}\gamma  + \left( {\frac{{25302017977}}{{4161798144}} - \frac{{451}}{{2048}}{\pi ^2}} \right)\eta  \\ 
  \left. { - \frac{{30913}}{{1835008}}{\eta ^2} + \frac{{235925}}{{1769472}}{\eta ^3} + \frac{{107}}{{280}}\log (2\theta )} \right\}{\theta ^6} \\ 
  \left. { + \left( { - \frac{{188516689}}{{433520640}} - \frac{{97765}}{{258048}}\eta  + \frac{{141769}}{{1290240}}{\eta ^2}} \right)\pi {\theta ^7}} \right] \\ 
\end{gathered}
\end{align*}
and phase is given by
\begin{align*}
\begin{gathered}
  \phi _{3.5}^{(T3)}(t) = \phi _{ref}^{(T3)} - \frac{1}{{\eta {\theta ^5}}}\left[ {1 + \left( {\frac{{3715}}{{8064}} + \frac{{55}}{{96}}\eta } \right){\theta ^2} - \frac{{3\pi }}{4}{\theta ^3} + \left( {\frac{{9275495}}{{14450688}} + \frac{{284875}}{{258048}}\eta  + \frac{{1855}}{{2048}}{\eta ^2}} \right){\theta ^4}} \right. \\ 
   + \left( {\frac{{38645}}{{21504}} - \frac{{65}}{{256}}\eta } \right)\ln \left( {\frac{\theta }{{{\theta _{lso}}}}} \right)\pi {\theta ^5} \\ 
   + \left\{ {\frac{{831032450749357}}{{57682522275840}}} \right. - \frac{{53}}{{40}}{\pi ^2} + \frac{{107}}{{56}}\gamma  + \left( { - \frac{{126510089885}}{{4161798144}} + \frac{{2255}}{{2048}}{\pi ^2}} \right)\eta  \\ 
  \left. { + \frac{{154565}}{{1835008}}{\eta ^2} - \frac{{1179625}}{{1769472}}{\eta ^3} - \frac{{107}}{{56}}\log (2\theta )} \right\}{\theta ^6} \\ 
  \left. { + \left( { - \frac{{188516689}}{{173408256}} + \frac{{488825}}{{516096}}\eta  + \frac{{141769}}{{516096}}{\eta ^2}} \right)\pi {\theta ^7}} \right] \\ 
\end{gathered}
\end{align*}
where
\begin{align*}
    \theta  = {\left[ {\eta ({t_{ref}} - t)/(5M)} \right]^{ - 1/8}}
\end{align*}
and the complete waveform is given by
\begin{align*}
    {h_A}(t) = Cv_A^2\sin \left[ {2{\phi _A}(t)} \right]
\end{align*}
The paper also includes the EOB (effective one body) model in its comparison.
\section{Our Work}
\subsection{Inspiral}
Using the 2PN frequency sweep equation \cite{poisson1995gravitational}
\begin{align*}
\frac{{dF}}{{dt}} = \frac{{96}}{{5\pi {M_{ch}}}}{(\pi {M_{ch}}F)^{11/3}}\left[ {1 - \left( {\frac{{743}}{{336}} + \frac{{11}}{4}\eta } \right){{(\pi MF)}^{2/3}}} \right.\\
 + (4\pi  - \beta )(\pi MF)\\
\left. { + \left( {\frac{{34103}}{{18144}} + \frac{{13661}}{{2016}}\eta  + \frac{{59}}{{18}}{\eta ^2} + \sigma } \right){{(\pi MF)}^{4/3}}} \right]
\end{align*}
We form the difference equation as
\begin{align*}
F(n) = F(n - 1) + {t_s}\frac{{96}}{{5\pi {M_{ch}}}}{(\pi {M_{ch}}F(n - 1))^{11/3}}\left[ {1 - \left( {\frac{{743}}{{336}} + \frac{{11}}{4}\eta } \right){{(\pi MF(n - 1))}^{2/3}}} \right.\\
 + (4\pi  - \beta )(\pi MF(n - 1))\\
\left. { + \left( {\frac{{34103}}{{18144}} + \frac{{13661}}{{2016}}\eta  + \frac{{59}}{{18}}{\eta ^2} + \sigma } \right){{(\pi MF(n - 1))}^{4/3}}} \right]
\end{align*}
using Euler's method, where
\begin{enumerate}
\item  $F(i)$ is the $i^{th}$ sample of the gravitational wave frequency
\item  $F(1)$ is the first sample (assumed as 10Hz)
\item $t_s$ is the sampling time given by $f_s = 1 / t_s$ where $f_s$ is the sampling frequency taken as 16384 Hz for LIGO.
\item  $M_{c} {}_{h} $ is the chirp mass of the binary black hole system calculated by
\[M_{ch} =\eta ^{3/5} M\] 
$\eta $ is the symmetric mass ratio calculated as
\[\eta =\mu /M\] 
where
\[\mu =m_{1} m_{2} /(m_{1} +m_{2} )\] 
and
\[M=m_{1} +m_{2} \] 
Here, $m_{1} =m_{2} =10M_{\odot } $ ,$1M_{\odot } =4.926\times 10^{-6} \sec $ in geometrized units \cite{cutler1994gravitational}.
\item  $\beta $ and $\sigma $ are the spin-orbit and spin-spin parameter respectively both considered to be zero for non-spinning case.
\end{enumerate}
Now, having already defined the initial value we need a cut-off frequency for the model which is given by
\[F_{ISCO} =\frac{6^{-3/2} }{\pi M} \] 
Which is obtained from the relation \cite{poisson1995gravitational}
\[\pi MF_{i} =(M/r_{i} )^{3/2} =6^{-3/2} \] 
where $r_{i} =6M$ is the Swarzschild radius of the innermost stable circular orbit (ISCO). Strain at the detector is given by \cite{balasubramanian1996gravitational,balasubramanian1996erratum}
\[h(t)=A[\pi f(t)]^{2/3} \cos [\phi (t)]\] 
Here f(t) is the instantaneous gravitational wave frequency, A is a constant that contains information about the distance from detector to the binary, the reduced and total mass of the system, and the antenna pattern of the detector. The 2PN phase can be expanded as \cite{balasubramanian1996gravitational,balasubramanian1996erratum}
\[\phi (t)=\phi _{0} (t)+\phi _{1} (t)+\phi _{1.5} (t)+\phi _{2} (t)\] 
Where
\[\phi (t)=\phi _{0} (t)=\frac{16\pi f_{a} \tau _{0} }{5} \left(1-\left(\frac{f}{f_{a} } \right)^{-5/3} \right)+\Phi \]
\[\phi _{1} (t)=4\pi f_{a} \tau _{1} \left[1-\left(\frac{f}{f_{a} } \right)^{-} {}^{1} \right]\] 
\[\phi _{1.5} (t)=-5\pi f_{a} \tau _{1.5} \left[1-\left(\frac{f}{f_{a} } \right)^{-} {}^{2/3} \right]\] 
\[\phi _{2} (t)=8\pi f_{a} \tau _{2} \left[1-\left(\frac{f}{f_{a} } \right)^{-} {}^{1/3} \right]\] 
here $\phi _{0} (t)$ is the dominant Newtonian part of the phase and $\phi _{n} $ is the nth order correction to it. Where f(t) is the instantaneous frequency of the gravitational wave given by \cite{balasubramanian1996gravitational,balasubramanian1996erratum}
\[\begin{array}{rcl} {t-t{}_{a} } & {=} & {\tau _{0} \left[1-\left(\frac{f}{f_{a} } \right)^{-8/3} \right]+\tau _{1} \left[1-\left(\frac{f}{f_{a} } \right)^{-2} \right]} \\ {} & {} & {+\tau _{1.5} \left[1-\left(\frac{f}{f_{a} } \right)^{-5/3} \right]+\tau _{2} \left[1-\left(\frac{f}{f_{a} } \right)^{-4/3} \right]} \end{array}\] 
And $f_{a} $  \& $\Phi $ are the frequency and phase at $t=t_{a}$. Here $f_a$ is taken as 10Hz \cite{fritschel2002low}. Now the chirp time has four components in total \cite{balasubramanian1996gravitational,balasubramanian1996erratum}
\[\tau _{0} =\frac{5}{256} M_{c} {}_{h} {}^{-5/3} (\pi f_{a} )^{-8/3} \] 
\[\tau _{1} =\frac{5}{192\mu (\pi f_{a} )^{2} } \left(\frac{743}{336} +\frac{11}{4} \eta \right)\] 
\[\tau _{1.5} =\frac{1}{8\mu } \left(\frac{M}{\pi ^{2} f_{a}^{5} } \right)^{1/3} \] 
\[\tau _{2} =\frac{5}{128\mu } \left(\frac{M}{\pi ^{2} f_{a} {}^{2} } \right)^{2/3} \left(\frac{3058673}{1016004} +\frac{5429}{1008} \eta +\frac{617}{144} \eta ^{2} \right)\] 
Now, $t_{c} $ is the sum of arrival time and total chirp time \& $\Phi _{c} $ is a combination of $\Phi $ and various chirp times \cite{balasubramanian1996gravitational,balasubramanian1996erratum}
\[t_{c} =t_{a} +\tau _{0} +\tau _{1} -\tau _{1.5} +\tau _{2} \] 
\[\Phi _{c} =\Phi +\frac{16\pi f_{a} }{5} \tau _{0} +4\pi f_{a} \tau _{1} -5\pi f_{a} \tau _{1.5} +8\pi f_{a} \tau _{2} \] 
\subsection{Meger-Ringdown}
For the merger-ringdown, the frequency evolution $\omega(t)$ and waveform amplitude $A(t)$ is given by \cite{huerta2017complete}
\begin{align*}
    \begin{gathered}
  \omega (t) = {\omega _{QNM}}(1 - \hat f) \hfill \\
  {\omega _{QNM}} = 1 - 0.63{(1 - {{\hat s}_{fin}})^{0.3}} \hfill \\
  A(t) = \frac{{{A_0}}}{{\omega (t)}}\left[ {\frac{{\left| {\dot {\hat f}} \right|}}{{1 + \alpha ({{\hat f}^2} - {{\hat f}^4})}}} \right] \hfill \\ 
\end{gathered}
\end{align*}
where ${{\hat s}_{fin}}$ is the spin of the black hole remnant. Also \cite{huerta2017complete}
\begin{align*}
    \begin{gathered}
  \hat f = \frac{c}{2}{\left( {1 + \frac{1}{\kappa }} \right)^{1 + \kappa }}\left[ {1 - {{\left( {1 + \frac{1}{\kappa }{e^{ - 2t/b}}} \right)}^{ - \kappa }}} \right] \hfill \\
  {{\hat s}_{fin}} = 2\sqrt 3 \eta  - \frac{{390}}{{79}}{\eta ^2} + \frac{{2379}}{{287}}{\eta ^3} - \frac{{4621}}{{276}}{\eta ^4} \hfill \\ 
\end{gathered}
\end{align*}
and the constants \cite{huerta2017complete}
\begin{align*}
    \begin{gathered}
  b(\eta ) = \frac{{16014}}{{979}} - \frac{{29132}}{{1343}}{\eta ^2} \hfill \\
  c(\eta ) = \frac{{206}}{{903}} + \frac{{180}}{{1141}}\sqrt \eta   + \frac{{424}}{{1205}}\frac{{{\eta ^2}}}{{\log (\eta )}} \hfill \\
  \kappa (\eta ) = \frac{{713}}{{1056}} - \frac{{23}}{{193}}\eta  \hfill \\
  \alpha (\eta ) = \frac{1}{{{Q^2}({{\hat s}_{fin}})}}\left( {\frac{{16313}}{{562}} + \frac{{21345}}{{124}}\eta } \right) \hfill \\
  Q({{\hat s}_{fin}}) = \frac{2}{{{{(1 - {{\hat s}_{fin}})}^{0.45}}}} \hfill \\ 
\end{gathered}
\end{align*}
The merger waveform can be obtained by \cite{huerta2017complete}
\begin{align*}
    \begin{gathered}
  {h^{merger}}(t) = {h_ + }^{merger}(t) - i{h_ \times }^{merger}(t) = A(t){e^{ - {\Phi _{gIRS}}(t)}} \hfill \\
  {\Phi _{gIRS}}(t) = \int_{{t_0}}^t {\omega (t)dt}  \hfill \\ 
\end{gathered}
\end{align*}
We can obtain $\Phi _{gIRS}(t)$ by using the following integrated solution
\begin{align*}
    \int {\omega (t)dt = {\omega _{QNM}}\left\{ {t - z\left[ {t - \left( {\frac{{b\kappa }}{2}{y^{\frac{k}{{k - 1}}}} + \frac{{b\kappa }}{{2\left( {\kappa  + 1} \right)}}{y^{\frac{{k + 1}}{{k - 1}}}}} \right)} \right]} \right\}}
\end{align*}
where
\begin{align*}
\begin{gathered}
  z = \frac{c}{2}{\left( {1 + \frac{1}{\kappa }} \right)^{1 + \kappa }} \hfill \\
  y = {\left( {1 - \frac{1}{{\kappa x}}} \right)^{\kappa  - 1}} \hfill \\
  x = {e^{2t/b}} + \frac{1}{\kappa } \hfill \\
\end{gathered}
\end{align*}
and $\dot{ \hat f}$ is given as
\begin{align*}
\dot {\hat f} =  - \frac{c}{b}{e^{ - 2t/b}}{\left( {1 + \frac{1}{\kappa }} \right)^{1 + \kappa }}{\left( {1 + \frac{{{e^{ - 2t/b}}}}{\kappa }} \right)^{ - \kappa  - 1}}
\end{align*}
Detailed derivations of $\Phi _{gIRS}(t)$ and $\dot{ \hat f}$ can be found at the end of the report in the supplementary materials. To combine the inspiral model with the merger model refer to \cite{huerta2017complete}.
\subsection{Simulation Results}
\begin{figure}[H]
    \centering
    \includegraphics[width=1\textwidth]{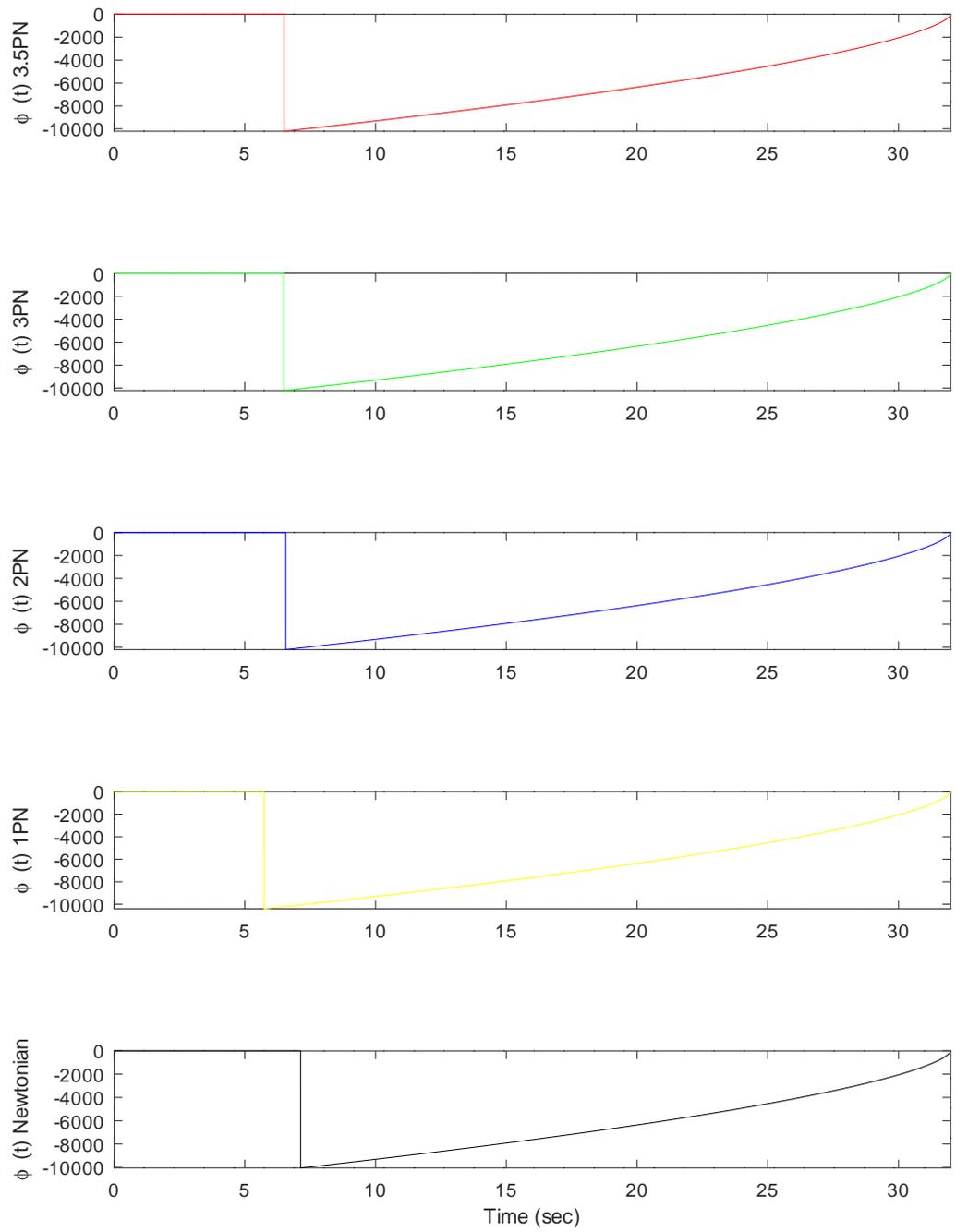}
    \caption{Phase Comparison}
    \label{fig:phase1}
\end{figure}

\begin{figure}[H]
    \centering
    \includegraphics[width=1\textwidth]{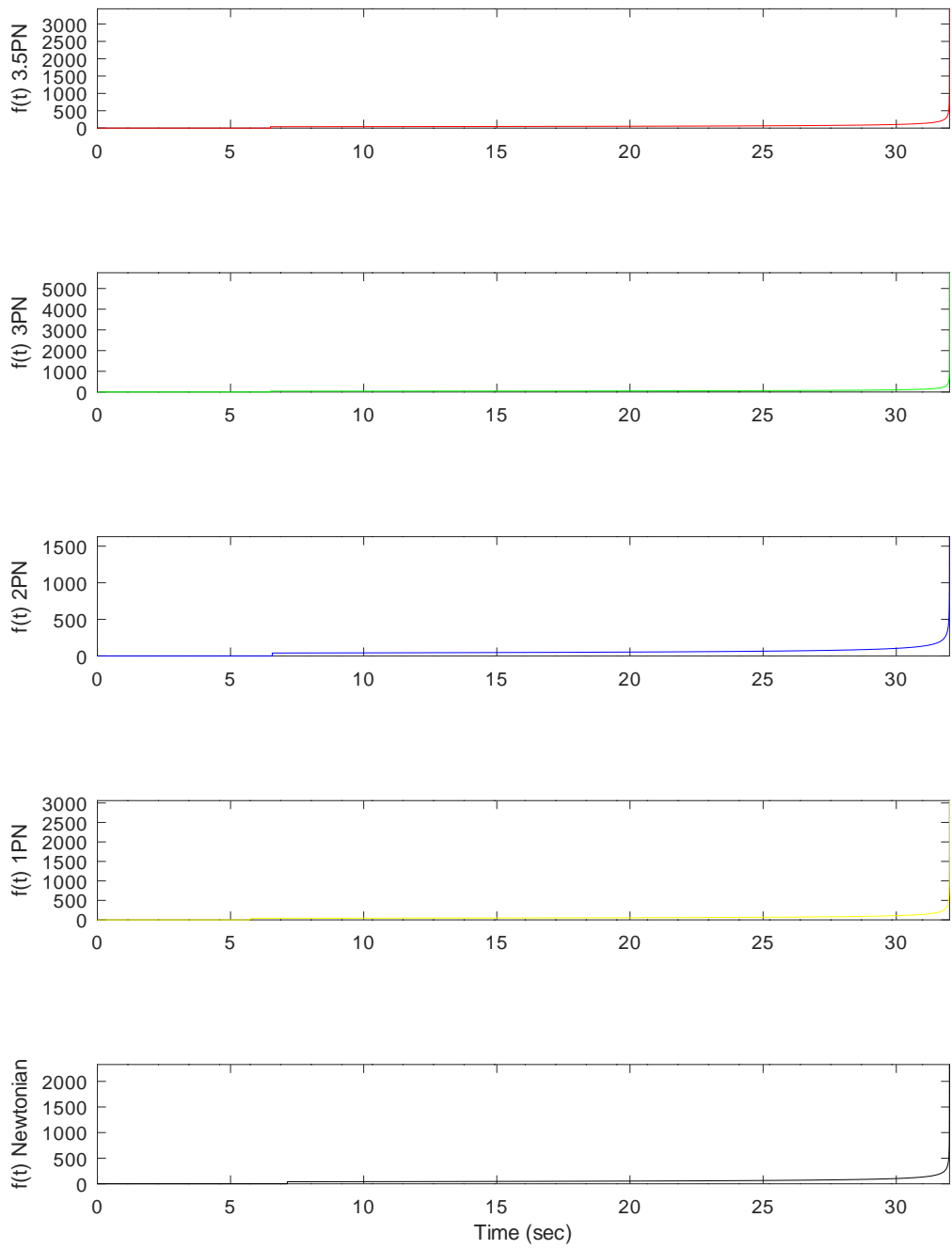}
    \caption{Frequency Comparison}
    \label{fig:freq1}
\end{figure}

\begin{figure}[H]
    \centering
    \includegraphics[width=1\textwidth]{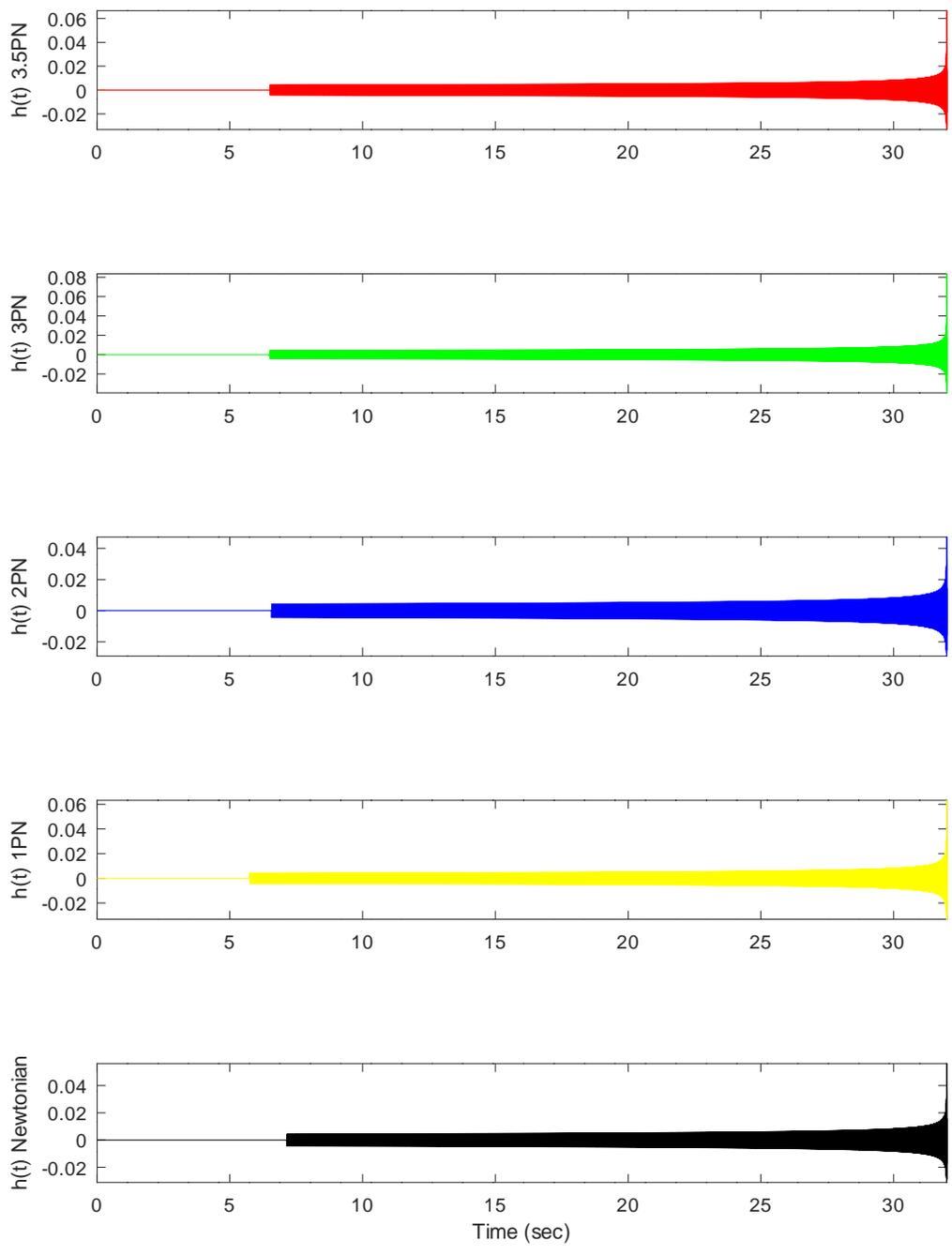}
    \caption{Waveform Comparison}
    \label{fig:wave1}
\end{figure}

The following results are for the inspiral only. They provide a comparison between the various post-Newtonian approximations with respect to the frequency, and how this impacts the end result. Fig.\ref{fig:phase1} gives a comparison of the phases, Fig.\ref{fig:freq1} gives a comparison of the frequencies, and Fig.\ref{fig:wave1} gives a comparison of the waveforms, of various post-Newtonian approximations.
\section{\textit{\textbf{Supplementary Materials}}}
\subsection{Deriving $\Phi _{gIRS}$}
To derive $\Phi _{gIRS}(t)$, we need to integrate $\omega (t)$ which is given by
\begin{align*}
    \omega (t) = {\omega _{QNM}}(1 - \hat f)
\end{align*}
where $\omega _{QNM}$ is a constant. Expanding $\omega (t)$ we get
\begin{align*}
    \omega (t) = {\omega _{QNM}} - {\omega _{QNM}}{\hat f}
\end{align*}
Integrating the above equation we obtain
\begin{align*}
    \int {\omega (t)dt = {\omega _{QNM}}t - {\omega _{QNM}}\int {\hat fdt} }
\end{align*}
Now
\begin{align*}
    \hat f = \frac{c}{2}{\left( {1 + \frac{1}{\kappa }} \right)^{1 + \kappa }}\left[ {1 - {{\left( {1 + \frac{1}{\kappa }{e^{ - 2t/b}}} \right)}^{ - \kappa }}} \right] 
\end{align*}
Lets assume the constant
\begin{align*}
    z = \frac{c}{2}{\left( {1 + \frac{1}{\kappa }} \right)^{1 + \kappa }}
\end{align*}
So we now have
\begin{align*}
    \hat f = {z}\left[ {1 - {{\left( {1 + \frac{1}{\kappa }{e^{ - 2t/b}}} \right)}^{ - \kappa }}} \right]
\end{align*}
Expanding which we get
\begin{align*}
    \hat f = \left[ {z - z{{\left( {1 + \frac{1}{\kappa }{e^{ - 2t/b}}} \right)}^{ - \kappa }}} \right]
\end{align*}
Hence
\begin{align*}
    \int {\hat fdt} = \left[ {zt - z\int{{\left( {1 + \frac{1}{\kappa }{e^{ - 2t/b}}} \right)}^{ - \kappa }}}dt \right]
\end{align*}
Now we need to solve the following
\begin{align*}
    \int{{\left( {1 + \frac{1}{\kappa }{e^{ - 2t/b}}} \right)}^{ - \kappa }}dt
\end{align*}
which we do as
\begin{align*}
   \begin{gathered}
  \int {{{\left( {1 + \frac{{{e^{ - 2t/b}}}}{\kappa }} \right)}^{ - \kappa }}dt}  \\ 
   = \int {\frac{1}{{{{\left( {1 + \frac{{{e^{ - 2t/b}}}}{\kappa }} \right)}^\kappa }}}} dt \\ 
   = \int {\frac{1}{{{{\left( {1 + \frac{1}{{\kappa {e^{2t/b}}}}} \right)}^\kappa }}}} dt \\ 
   = \int {\frac{1}{{{{\left( {\frac{{\kappa {e^{2t/b}} + 1}}{{\kappa {e^{2t/b}}}}} \right)}^\kappa }}}} dt \\ 
   = \int {{{\left( {\frac{{\kappa {e^{2t/b}} + 1}}{{\kappa {e^{2t/b}}}}} \right)}^\kappa }dt}  \\ 
   = \int {\frac{{{e^{2t\kappa /b}}}}{{{{\left( {{e^{2t/b}} + \frac{1}{\kappa }} \right)}^\kappa }}}} dt \\ 
\end{gathered}
\end{align*}
Now lets assume the following
\begin{align*}
    \begin{gathered}
  \frac{1}{\kappa } + {e^{2t/b}} = x \\ 
  \frac{2}{b}{e^{2t/b}}dt = dx \\ 
  dt = \frac{b}{2}\frac{1}{{\left( {x - \frac{1}{\kappa }} \right)}}dx \\ 
\end{gathered}
\end{align*}
So the integration becomes
\begin{align*}
    \begin{gathered}
  \frac{b}{2}\int {\frac{{{{\left( {x - \frac{1}{\kappa }} \right)}^{\kappa  - 1}}}}{{{x^\kappa }}}dx}  \\ 
   = \frac{b}{2}\int {\frac{{\frac{{{x^{\kappa  - 1}}}}{{{x^{\kappa  - 1}}}}{{\left( {x - \frac{1}{\kappa }} \right)}^{\kappa  - 1}}}}{{{x^\kappa }}}dx}  \\ 
   = \frac{b}{2}\int {\frac{{{{\left( {1 - \frac{1}{{\kappa x}}} \right)}^{\kappa  - 1}}}}{x}dx}  \\ 
   = \frac{{b\kappa }}{2}\int {\frac{1}{{\kappa x}}{{\left( {1 - \frac{1}{{\kappa x}}} \right)}^{\kappa  - 1}}dx}  \\ 
\end{gathered}
\end{align*}
Now assuming the following
\begin{align*}
    \begin{gathered}
  1 - \frac{1}{{\kappa x}} = {y^{\frac{1}{{\kappa  - 1}}}} \\ 
  dx = \frac{{\kappa {{\left( {\kappa \left( {1 - {y^{\frac{1}{{\kappa  - 1}}}}} \right)} \right)}^2}{y^{\frac{{2 - \kappa }}{{\kappa  - 1}}}}}}{{\kappa  - 1}}dy \\ 
\end{gathered}
\end{align*}
The integration becomes
\begin{align*}
    \begin{gathered}
  \frac{{b\kappa }}{2}\int {\frac{{y\kappa \left( {1 - {y^{\frac{1}{{\kappa  - 1}}}}} \right){y^{\frac{{2 - \kappa }}{{\kappa  - 1}}}}}}{{\kappa  - 1}}dy}  \hfill \\
   = \frac{{b{\kappa ^2}}}{{2\left( {\kappa  - 1} \right)}}\int {\left( {1 - {y^{\frac{1}{{\kappa  - 1}}}}} \right){y^{\frac{1}{{\kappa  - 1}}}}dy}  \hfill \\
   = \frac{{b{\kappa ^2}}}{{2\left( {\kappa  - 1} \right)}}\int {\left( {{y^{\frac{1}{{\kappa  - 1}}}} - {y^{\frac{2}{{\kappa  - 1}}}}} \right)dy}  \hfill \\
   = \frac{{b{\kappa ^2}}}{{2\left( {\kappa  - 1} \right)}}\left[ {\frac{{{y^{\frac{1}{{\kappa  - 1}} + 1}}}}{{\frac{1}{{\kappa  - 1}} + 1}} - \frac{{{y^{\frac{2}{{\kappa  - 1}} + 1}}}}{{\frac{2}{{\kappa  - 1}} + 1}}} \right] \hfill \\
   = \frac{{b\kappa }}{2}{y^{\frac{k}{{\kappa  - 1}}}} + \frac{{b{\kappa ^2}}}{{2\left( {\kappa  + 1} \right)}}{y^{\frac{{k + 1}}{{\kappa  - 1}}}} \hfill \\ 
\end{gathered} 
\end{align*}
where
\begin{align*}
    \begin{gathered}
  y = {\left( {1 - \frac{1}{{\kappa x}}} \right)^{\kappa  - 1}} \hfill \\
  x = {e^{2t/b}} + \frac{1}{\kappa } \hfill \\ 
\end{gathered}
\end{align*}
So the final equation becomes
\begin{align*}
    \int {\omega (t)dt = {\omega _{QNM}}\left\{ {t - z\left[ {t - \left( {\frac{{b\kappa }}{2}{y^{\frac{k}{{k - 1}}}} + \frac{{b\kappa }}{{2\left( {\kappa  + 1} \right)}}{y^{\frac{{k + 1}}{{k - 1}}}}} \right)} \right]} \right\}}
\end{align*}
with $x$,$y$ and $z$ having already been defined.
\subsection{Deriving $\dot {\hat f}$}
To derive $\dot {\hat f}$ we need to differentiate ${\hat f}$
\begin{align*}
    \hat f = \frac{c}{2}{\left( {1 + \frac{1}{\kappa }} \right)^{1 + \kappa }}\left[ {1 - {{\left( {1 + \frac{1}{\kappa }{e^{ - 2t/b}}} \right)}^{ - \kappa }}} \right] 
\end{align*}
Lets assume the constant
\begin{align*}
    z = \frac{c}{2}{\left( {1 + \frac{1}{\kappa }} \right)^{1 + \kappa }}
\end{align*}
So we now have
\begin{align*}
    \hat f = {z}\left[ {1 - {{\left( {1 + \frac{1}{\kappa }{e^{ - 2t/b}}} \right)}^{ - \kappa }}} \right]
\end{align*}
Expanding which we get
\begin{align*}
    \hat f = \left[ {z - z{{\left( {1 + \frac{1}{\kappa }{e^{ - 2t/b}}} \right)}^{ - \kappa }}} \right]
\end{align*}
Now differentiating the above using chain rule we get
\begin{align*}
    \begin{gathered}
  \dot \hat f =  - z\left( { - \kappa {{\left( {1 + \frac{{{e^{ - 2t/b}}}}{\kappa }} \right)}^{ - \kappa  - 1}}} \right)\left( {\frac{{{e^{ - 2t/b}}}}{\kappa }\left( {\frac{{ - 2}}{b}} \right)} \right) \\ 
   =  - \frac{{2z{e^{ - 2t/b}}}}{b}{\left( {1 + \frac{{{e^{ - 2t/b}}}}{\kappa }} \right)^{ - \kappa  - 1}} \\ 
\end{gathered}
\end{align*}
and putting the value of $z$ finally
\begin{align*}
    \dot \hat f =  - \frac{{c{e^{ - 2t/b}}}}{b}{\left( {1 + \frac{1}{\kappa }} \right)^{1 + \kappa }}{\left( {1 + \frac{{{e^{ - 2t/b}}}}{\kappa }} \right)^{ - \kappa  - 1}}
\end{align*}
\bibliographystyle{unsrt}
\bibliography{sample}
\appendix
\setcounter{secnumdepth}{0}
\end{document}